\documentclass{ws-ijmpaMod}
\usepackage{color}
\newcommand{\Mpl}{M_{\rm pl}}
\newcommand{\M}{{\cal M}}

\renewcommand{\d}{{\rm d}}
\newcommand{\ep}{\epsilon}
\newcommand{\fnl}{f_{\mathrm{NL}}}
\newcommand{\be}{\begin{equation}}
\newcommand{\ee}{\end{equation}}
\newcommand{\bea}{\begin{eqnarray}}
\newcommand{\eea}{\end{eqnarray}}

\begin{document}

\markboth{Joseph Elliston, David Mulryne, David Seery and Reza Tavakol}{Evolution of Non-Gaussianity in Multi-scalar Field Models}

%
\catchline{}{}{}{}{}
%

\title{EVOLUTION OF NON-GAUSSIANITY IN MULTI-SCALAR FIELD MODELS}

\author{Joseph Elliston\footnote{j.elliston@qmul.ac.uk,
$^{\dag}$d.mulryne@qmul.ac.uk, $^{\ddag}$d.seery@sussex.ac.uk,
$^{\Diamond}$r.tavakol@qmul.ac.uk} $^{1}$,
David Mulryne$^{\dag 1}$, David Seery$^{\ddag 2}$ and
Reza Tavakol$^{\Diamond 1}$}

\address{$^{1}$Astronomy Unit,
Queen Mary, University of London,
Mile End Road, London, E1 4NS, UK.\\
$^{2}$Astronomy Centre, University of Sussex, Falmer, Brighton, BN1 9QH, UK.}

\maketitle


\begin{abstract}
We study the evolution of non-Gaussianity in
multiple-field inflationary models,
focusing on three fundamental questions:
(a) How is the sign and peak magnitude
of the non-linearity parameter $\fnl$
related to generic features in the inflationary
potential?
(b) How sensitive is $\fnl$
to the process by which an adiabatic limit is reached, where the curvature
perturbation becomes
conserved? (c) For a given model,
what is the appropriate tool -- analytic or numerical --
to calculate $\fnl$ at the adiabatic limit?
We summarise recent results obtained by the authors
and further elucidate them by considering an inflection point model.
\end{abstract}

\ccode{PACS numbers: 98.80.-k, 98.80.Cq}

\section{Introduction}

An important feature of canonical single field models is that
the curvature perturbation produced at horizon crossing is
conserved,\cite{Lyth,zetaAndZetaCon}
with statistics indistinguishable from Gaussian.\cite{Maldacena:2002vr}
In multiple field models, on the other hand,
isocurvature modes may also
be produced, and can subsequently source the evolution of the curvature
perturbation as the field space path
curves.\cite{GarciaBellido:1995qq,Gordon:2000hv}
The curvature perturbation and its statistics
(such as the power spectrum and non-Gaussianity\cite{Vernizzi:2006ve,otherMethods}) can therefore continuously
evolve after horizon crossing, leading to far richer behaviour than that
allowed in single field models.
Here we focus on non-Gaussianity, and in particular the nonlinearity
parameter, $\fnl$, though many of our general conclusions
extend to other observables as well.

In canonical multi-field models,
the non-Gaussianity present at
horizon crossing is negligible.\cite{Seery:2005gb}
In this setting, however, it is possible for $\fnl$ to evolve
to large values.\cite{hybrid,Byrnes:2008wi,Kim:2010ud,Peterson:2010mv}
To calculate the observationally relevant value of non-Gaussianity,
in principle one has to follow the evolution until the
time of last scattering, where the Cosmic Microwave Background (CMB) was imprinted.
Given our present ignorance about the detailed physics of the early universe,
this would not be possible in practice. In many
models, however, a regime is reached during the evolution
long before this time -- the so-called {\it adiabatic limit} --
where all the isocurvature modes decay, and the
curvature perturbation becomes conserved.

There are different ways in which such a limit could be reached, and this
has consequences both for the possible values of the observable parameters, such as
$\fnl$, and for the techniques which can be reliably employed to calculate
observables, i.e. whether analytic methods will suffice, or
numerical methods are necessary.
A useful classification of models
according to when the adiabatic limit is attained is:
\begin{itemize}
\item{\it Models in which an adiabatic limit is reached `naturally'
by convergence into a `focusing region' of the
potential. This can be further
sub-divided into cases where the convergence occurs
during slow-roll, and cases in which convergence
occurs only after the slow-roll approximation fails.}
\vskip 0.1in
\item {\it Models where an adiabatic limit is reached abruptly
due to an additional degree of freedom, such as a
waterfall field being destabilised.}
\vskip 0.1in
\item {\it Models for which no focusing
region in the inflationary potential exists, and
an adiabatic limit can be reached only by embedding the inflationary
model into a larger scenario, perhaps one which includes perturbative
reheating.}
\end{itemize}

Recently an extended study of these possibilities
was undertaken by the authors in the context of
of multi-field models of inflation, capable of producing
large non-Gaussianities.\cite{EMST-11}
Here we give a summary of those results and
further elucidate them by considering an inflection point model, which is of
the first type,
and which can produce a large positive or negative value of $\fnl$ at the adiabatic limit.

The structure of the paper is as follows. In
\S \ref{background} we give a summary of the background theory which is
used to formulate analytic expressions for observables,
discussed in \S \ref{sec:analytics}, together with the
conditions needed for the non-Gaussianity to be large at
the adiabatic limit when it is reached naturally.
In \S \ref{sec:largeNG} we briefly discuss features in the potential
which generate large transitory non-Gaussianities during the 
super-horizon evolution, which may be relevant for the
final observable value at the adiabatic limit if this limit
is reached abruptly. Finally, we discuss the usefulness of our results in
\S \ref{sec:models} when applied to
specific models, and demonstrate this by considering a new example. We
conclude in \S \ref{sec:conclusions}.

\section{Background} \label{background}
We consider inflation driven by multiple
canonical scalar fields $\phi_i$ with $i = 1, 2, ... ,\, \M$,
self-interacting through a potential $W(\phi_1 , \phi_2 , ...)$.
Defining $W_{,i} = {\partial W}/{\partial \phi_i}$
the scalar equations of motion are
\be
\label{eq:eoms}
\ddot \phi_i + 3H \dot \phi_i + W_{,i} = 0\,,
\ee
where $H$ is the Hubble rate, given by the associated Friedmann equation
$3 \Mpl^2 H^2 = W + \sum_{i} \dot \phi_i^2/2 $.
We define slow-roll parameters as
\be
\epsilon_i = \frac{\Mpl^2}{2} \left (\frac{{W_{,i}}}{W} \right)^2,~~~
\epsilon = \sum_{i=1}^\M \epsilon_i, ~~~
\eta_{ij} = {\Mpl^2} \left (\frac{W_{,ij}}{W} \right )\,,
\ee
such that for inflation we require $\epsilon < 1$.
The `slow-roll limit' is given by $\epsilon \ll 1$,
during which the fields' kinetic energy may be
neglected, the decaying modes discarded, and the
field equations well approximated by
\be
\label{eq:sr_eoms}
3H \dot \phi_i + W_{,i} = 0, ~~~~3 \Mpl^2 H^2 = W.
\ee

Primordial cosmological perturbations are commonly characterised in terms of
the curvature perturbation on uniform density spatial hypersurfaces,
denoted by $\zeta$. An important feature of $\zeta$ is
that for adiabatic perturbations it is conserved on large
scales,\cite{zetaAndZetaCon} at the linear order and even beyond.
In multi-field models, on the other hand,
$\zeta$ can evolve due to the presence of isocurvature
modes, and this may result in the production of large non-Gaussianities.
Deviation of the three-point function from zero
is commonly measured in terms of the dimensionless
parameter $\fnl$
\be
\label{fnl}
\fnl=\frac56\frac{k_1^3k_2^3k_3^3}{k_1^3+k_2^3+k_3^3}
\frac{{\cal B}_{\zeta}(k_1,k_2,k_3)}{4\pi^4 {\cal P}_{\zeta}^2},
\ee
where ${\cal P}_\zeta$ is the power spectrum and ${\cal B}_\zeta$ the bispectrum, given respectively by
\begin{eqnarray} \label{PS}
\langle\zeta_{\mathbf k_1}\zeta_{\mathbf k_2}\rangle &\equiv& (2\pi)^3
\delta^3 ({\mathbf k_1} + {\mathbf k_2}) \frac{2\pi^2}{{k_1}^3}{\cal P}_{\zeta}(k_1) \, , \\
\langle\zeta_{\mathbf k_1}\,\zeta_{\mathbf k_2}\,
\zeta_{\mathbf k_3}\rangle &\equiv& (2\pi)^3 \delta^3 ( {{\mathbf k_1}+{\mathbf k_2}+
{\mathbf k_3}}) {\cal B}_\zeta( k_1,k_2,k_3) \,.
\end{eqnarray}

A common technique for calculating $\zeta$ and its statistics, including
$\fnl$, is the $\delta N$ formalism,\cite{Starobinsky:1986fxa,Sasaki:1995aw,Lyth:2005fi}
based on the separate universe
approach to perturbation theory.\cite{Lyth,Wands:2000dp}
In this approach,
spatial gradients are neglected on scales greater than the horizon size,
and each spatial point is assumed to evolve as a separate FRW universe.
In phase space, this can be represented by
a bundle of trajectories, each evolving along an independent
path from perturbatively different initial conditions.
The variables which parametrise this phase space are the scalar fields
$\{\phi_i,\dot{\phi_i} \}$, as well as any radiation or matter species
that may be present. The idea of associating inflationary perturbations
with trajectories has a long history.\cite{Lyth,Starobinsky:1986fxa,GarciaBellido:1995qq,trajectories}

In this picture, choosing a different spatial slicing of the universe
corresponds to taking a different cross section of the bundle in phase space.

Choosing a flat initial slicing at $t=t^*$, and a later
uniform density (constant $H$) slicing at $t=t_c$,
then $\zeta$ on the final slicing can be equated with
the difference in the number of e-folds,
as measured along different trajectories in the bundle,
$\zeta = \delta N$.

In general, as the bundle evolves in the field space, so will $\delta N$
and its statistics, requiring the dynamics to be followed indefinitely.
If, however, the trajectories converge to a line parametrised by a
single variable, i.e. the adiabatic limit,
$\zeta$ becomes conserved.
Moreover, in this limit each value of the
Hubble rate corresponds to a single combination of the field
or fluid content,
implying that derivatives of final quantities on $c$, with respect to
changes in the initial conditions, tend to zero as the adiabatic limit
is reached. In particular, one finds $\partial \phi_i^c/\partial \phi_k^*\to 0$.

Restricting our attention to purely scalar field dynamics, a common
example of how an adiabatic limit is reached is for the
trajectories to evolve into a focusing region of the potential, such
as a potential valley, possibly terminating in a minimum.
For convergence into a valley, the mass-squared matrix
associated with perturbations orthogonal
to the direction to which the trajectories are converging
should have large and positive eigenvalues.
Taking the smallest eigenvalue to be of magnitude $\sim m_\perp$,
one typically expects a decay of the field derivatives
$\partial \phi_i^c/\partial \phi_k^*$ at least as fast as
$e^{-(m_\perp N) / (3H)}$ (see our recent work\cite{EMST-11} 
for a detailed discussion). As 
was mentioned in the introduction, however, an adiabatic
limit could be reached in other ways. It may turn out that no
focusing region in the potential is available, in which case the decay
of the fields into radiation may need
be considered in order for the model to make unambiguous predictions.
Alternatively, an adiabatic limit could occur due to a sudden transition,
such as a waterfall field being destabilised.

In any case, before we can consider the fate of observables at the
adiabatic limit, it is first necessary to have calculable
expressions for these quantities, which could be obtained
using the $\delta N$ formalism discussed above.
We recall that during slow-roll inflation field velocities are
functions of field positions. Taking this to be a good
approximation at horizon crossing, the subsequent number of e-folds
undergone by any `separate universe' is then
a function purely of the initial field values, $N(\phi_1^*,...,\phi_{\cal M}^*)$,
even if the evolution subsequently evolves
away from slow roll. Taking $t^*$ as a time shortly
after observable scales left the horizon, therefore, a Taylor expansion
\be \label{eq:deltaN}
\zeta \equiv \delta N = \sum_{i} N_{,i} \delta \phi_i^*
+ \frac{1}{2} \sum_{ij} N_{,ij} \delta \phi_i^* \delta \phi_j^* + \dots
\ee
can be made, where $N$ is the number of e-folds
from $*$ to $c$, a subscript $i$ represents
a derivative with respect to $\phi^*_i$, and $\delta \phi_k^*$
are the field fluctuations on the flat hyper-surface at horizon crossing.
Such a Taylor series
allows the properties of a bundle of trajectories to be parametrised
by just a few numbers, namely the derivatives of $N$ about some
typical member of the bundle.
Moreover, Eq. \eqref{eq:deltaN} allows various statistics to be estimated.
In particular $\fnl$ is given 
by\cite{Lyth:2005fi}
\be
\label{eq:fnl}
\fnl = \frac{5}{6} \frac{ \sum_{i,j} N_{,i} N_{,j} N_{,ij} }{\left( \sum_{i} N_{,i}^2 \right) ^2}.
\ee
\section{Analytic Schemes and $\fnl$ at a
`Natural' Adiabatic Limit} \label{sec:analytics}
To analytically evaluate the non-linearity parameter $\fnl$ from
Eq. (\ref{eq:fnl}) we must calculate $N_{,i}$ and $N_{,ij}$ at time $t^c$,
or when they become constant at the adiabatic limit.
Currently, analytic calculation is only possible when the
slow-roll equations of motion, Eq. (\ref{eq:sr_eoms}), are
a good approximation. This means that if $\zeta$ is still
evolving at the end of inflation, we cannot analytically follow observable
quantities, and numerical simulations will become essential\footnote{We are not considering models such as the curvaton, where
approximate analytic formula can be derived for regimes after the end of inflation by modelling the curvaton field as a fluid.}.
Moreover, calculations require a special `separable'
form for the potential\cite{GarciaBellido:1995qq}. Vernizzi \& Wands\cite{Vernizzi:2006ve}
and later Battefeld \& Easther\cite{Battefeld:2006sz}
studied sum-separable models,
$W = \sum_i V_i(\phi_i)$, deriving expressions for these coefficients
and for $\fnl$. Similar
techniques were used by Choi {\it et al.}\cite{Choi:2007su} for models of
product-separable form $W = \Pi_i V_i(\phi_i)$, and
recently Wang\cite{Wang:2010si} generalised the study of sum-type
potentials to those of the form $W = (\sum_i V_i(\phi_i))^{1/A}$, where
$A$ is an arbitrary constant\footnote{The product-separable
\cite{Choi:2007su} and the generalised sum separable
potentials\cite{Wang:2010si} were explicitly only considered with two fields, but the
results are easily generalised to an arbitrary number.\cite{EMST-11}}
A summary of analytic expressions for
$\M$-field models of these forms
is given in our paper.\cite{EMST-11}

The analytic formulae follow by using
Eq.~\eqref{eq:sr_eoms} to write
$N$ as an integral over one of the fields $\phi_k$ as
$N=-\int_{\phi^*}^{\phi^c} \! \! W/(\Mpl^2 W_{,\phi_k}) \, \d \phi_k$.
In general, taking the derivative of this expression
yields three contributions, namely initial and final boundary terms and a path term.
For potentials for which analytic progress is possible, however,
the path terms are either absent, or the integrals
can be manipulated to make them so.
For potentials with product-separable forms one finds
\be \label{eq:ps_Ni}
N^{(k)}_{,i}=\left. \frac{V_k}{\Mpl^2 V_k'} \right|_* \delta_{ik} - \left. \frac{V_k}{\Mpl^2 V_k'} \right|_c \frac{\partial \phi_k^c}{\partial \phi_i^*} \,,
\ee
where the free index $k$ labels the $\M$ ways of writing $N_{,i}$, all of
which will lead to the same result once the $c$-dependent terms are
evaluated. The summation convention is
\emph{not used} anywhere in this paper. For potentials
of generalised sum separable
form one finds a similar expression
\be
\label{eq:ss_Ni}
N_{,i}= A\left[ \left.\frac{V_i}{\Mpl^2 V_i'} \right|_* -
\sum_{k=1}^\M \left. \frac{V_k}{\Mpl^2 V_k'} \right|_c
\frac{\partial \phi_k^c}{\partial \phi_i^*} \right]\,.
\ee
In both cases $N_{,ij}$ follows by differentiation.

The difficult step in deriving an
analytic expression for $N_{,i}$ is the calculation of
the $c$-dependent derivatives
in Eqs. \eqref{eq:ps_Ni}--\eqref{eq:ss_Ni}.
A case where analytic progress is much easier
occurs if the adiabatic limit is reached
during slow-roll inflation. If this
limit is reached by convergence into a valley,\cite{EMST-11} we expect
the $c$-dependent derivatives to
tend to zero at least as fast as the lightest isocurvature
mode decays.\cite{EMST-11} If sufficient time
is available for them to become negligible, this
greatly simplifies the expressions for $N_{,i}$. Indeed it is
often possible to set the entire $c$-dependent boundary
term to zero, and the expressions
become dependent only on the
values the fields took at horizon crossing. Where this simplification has been
used in the literature, it has been referred
to as the Horizon Crossing Approximation (HCA).\cite{HCA,Kim:2010ud}
Caution is needed, however,
since it is possible
that the coefficients $V_k / V_k'$ in front of the derivatives
may diverge as the
adiabatic limit approaches. This possibility
is discussed at length elsewhere,\cite{EMST-11} and is only
possible if the field, $\phi_k$, is completely orthogonal
to the final adiabatic direction (the valley bottom), \emph{and}
$V_k$ tends to a constant as $V_k'$ tends to
zero. In this case the $c$-dependent term will
tend to an unknown constant (unless the full calculation of the derivatives
is performed), rather than zero. In the case of sum--separable
potentials, however, such a possibility can be avoided since
we are free to reparametrise the
potential and associate the problematic constant with another field. In the
product-separable case we must simply pick the $k$th version of Eq.~\eqref{eq:ps_Ni},
associated with a field $\phi_k$ which is not orthogonal to the final
adiabatic direction. Once this procedure is followed, the second
derivatives of $N$ follow by differentiation\footnote{Formally,
one should first calculate the expression for $N_{,ij}$, and then allow
derivatives of $\phi^c_i$ to tend to zero. These two operations commute provided $(V_j /V'_j)'$
is finite at the adiabatic limit.}.
The technical details are presented in our paper.\cite{EMST-11}
Here we instead illustrate this point with the help of some examples.

Consider first a sum--separable potential of the form
$W = V_0(e^{\lambda \phi} + e^{\lambda \chi})$, which has a valley bottom
defined by the line $\phi = \chi$. Since neither field is orthogonal
to this final direction, both $V_\phi '$ and $V_\chi '$
remain non-zero as the adiabatic limit is asymptotically
approached, and both fields continue to evolve.
As the $c$-dependent derivatives vanish, therefore, so does
the entire $c$-dependent boundary term, leading to expressions
for $N_{,i}$ which depend only on initial conditions.
Alternatively, consider a potential of the form
$W = W_0 + \frac{1}{2} m^2_\phi \phi^2 + \frac{1}{2} m^2_\chi \chi^2$
with $m_\chi \gg m_\phi$.
In this case
the adiabatic limiting trajectory is the line defined by
$\chi = 0$. As convergence to this trajectory occurs, $V_\chi '$ vanishes and, if we
define $V_\chi = W_0 + \frac{1}{2} m^2_\chi \chi^2$, we
would arrive at an incorrect expression for the $N_{,i}$ by
na\"{i}vely setting the $c$-dependent terms to zero.
We are free, however,
to define instead $V_\chi = \frac{1}{2} m^2_\chi \chi^2$ and $V_\phi =W_0 + \frac{1}{2} m^2_\phi \phi^2$, and with this definition we can correctly set the $c$-dependent term to zero.

An example of a product-separable
potential with an convergent valley region is $W=W_0(1+ g \phi^2)\exp(-\lambda \chi^2)$.
In this case an adiabatic limiting trajectory
occurs when $\phi=0$, where the velocity of the $\phi$ field
tends to zero. Following the
discussion above, we select the way of writing $N_{,i}$ associated with
the field which is still evolving as the valley bottom is reached,
in this case $\chi$, and with this choice setting the $c$-dependent
term to zero
leads to an accurate asymptotic expression
for $N_{,i}$.

In this way any model with a separable potential can be analysed,
and one immediate question is of interest: \emph{Can $\fnl$ be large after
natural focusing?} Discarding the $c$-dependent terms one can
employ Eq.~(\ref{eq:ss_Ni}) together with
Eq. (\ref{eq:fnl}) to obtain for the product-separable potentials
the expression
\be
\label{eq:ps_fnl}
\frac{6}{5} \fnl = 2 \ep_k^* - \eta_{kk}^*.
\ee
where $\phi_k$ is the field still evolving at the adiabatic limit.
Thus we find that models with product-separable
potentials lead to slow-roll suppressed
values of non-Gaussianities if the adiabatic limit
is reached during slow-roll inflation.

The situation for sum separable
potentials is more complicated, but a relatively
simple picture emerges if we assume that $N_{,i}$ is much larger
for one field, $\phi$ say (or for just a few
fields as detailed by Kim {\it et al.}\cite{Kim:2010ud})
In this case $V_\phi/V_\phi'$ at horizon crossing is much greater than the analogous terms for the other fields,
and one finds
\be
\fnl=-\frac{5}{6}\Mpl^2 \eta_{\phi \phi} \frac{W}{V_\phi}\,.
\ee
Thus in single field inflation
$\fnl$ would be suppressed by the slow-roll parameter
$\eta$. However, in multiple-field models $V_\phi''/V_\phi$ need not be small
even when $\eta_{\phi \phi}$ is, if one or more of the other fields contribute
significantly to the energy density. The
condition for a large non-Gaussianity in this case is therefore that
the mass-squared of $\phi$ is much greater than magnitude of its potential
at horizon crossing (in addition
to the condition on $V_\phi/V_\phi'$). We note that the sign of $\fnl$
is opposite to that of $V_\phi''$.

\section{The Magnitude and Sign of Transient Non-Gaussianities}
\label{sec:largeNG}

Our primary interest is in the final
constant value of $\fnl$ after an adiabatic limit has been reached.
However, `transitory'
large values of $\fnl$ may also be relevant, since it is possible for
the inflationary dynamics to be interrupted, rapidly establishing an adiabatic limit. Examples may occur in models containing a waterfall field
In that case, an evolving but large $\fnl$ value
can be preserved at the adiabatic limit. The general
conditions for such a large evolving value of $\fnl$ were given by
Byrnes {\it et al.},\cite{Byrnes:2008wi} using
the full analytic expressions for separable potentials.\cite{Vernizzi:2006ve,Choi:2007su}
In our recent work\cite{EMST-11}
a different
point of view was adopted. We studied the
corresponding conditions which are
required for a large $\fnl$ to be produced by features
that commonly occur in multi-field potentials.
We developed an approximation, based on intuition
from the phase-space picture of inflationary
trajectories, which allows simple scalings to be derived
for the peak magnitudes of the transitory $\fnl$ as well as their expected signs.
Space does not permit us to detail the entire derivation, so here we briefly
summarise the conclusions found.

To date models studied in the literature, which are capable of
producing large transitory $\fnl$, possess two broad features in their potentials:
a ridge or a valley, with a large $\fnl$ produced as
the bundle falls from a ridge or begins its turn into the
bottom of a valley. A common feature present in both these cases is
the rotation of the bundle of trajectories which results
in sourcing the evolution of $\zeta$
from isocurvature modes. Moreover, both cases
lead to a highly non-linear dependence of $N$
on initial conditions in the early stages of the
turn. This is because one side of the bundle of trajectories begins to turn
before the other, leading to a temporary asymmetry in the
dynamics. In general this non-linear dependence is encoded in the $\delta N$
formalism through large values of the second derivatives $N_{, ij}$.

To demonstrate this behaviour more concretely,
we specialised to two--field ($\phi, \chi$) potentials
and considered generic ridge and valley potential forms obtained
by perturbative expansions about some position along a separatrix
or the bottom of a valley, respectively.

For the ridge, the relevent potential takes the form\cite{EMST-11}
$W = W_0 + g \phi - \frac{1}{2} m_\chi^2 \chi^2$, where $W_0$,
$g$ and $m_\chi$ are constants, and $\chi=0$ defines the ridge.
We assumed that
the initial field position was sufficiently close to $\chi=0$ so that
the evolution was initially almost entirely in the $\phi$ direction.
Employing the $\delta N$ formalism we then showed that
the evolution leads to a negative peak in $\fnl$
with its maximum value scaling as\cite{EMST-11}
\be
\label{eq:ridge_fnl}
\left. \fnl \right|_{\rm Max} \sim -0.3 \epsilon_*^{1/2} \frac{\Mpl}{\chi_*}.
\ee
This inverse scaling with $\chi^*$ was explicitly verified
using a number of potentials containing ridge features.\cite{EMST-11}
Two important points are worth noting here. Firstly
the sign of $\fnl$ resulting from a ridge feature is \emph{negative} and
secondly, a large value of $\fnl$ requires a high degree of fine tuning
of the form $\chi^* \to 0$.

In general the valley evolution is more complicated.
However, a similar picture emerges when a perturbative expansion of the form
$W = W_0 + \frac{1}{2} m_\phi^2 \phi^2 + \frac{1}{2} m_\chi^2 \chi^2$, is considered
where $W_0$, $m_\phi$ and $m_\chi$ are constants.\cite{EMST-11}
In this case we assumed $m_\phi \gg m_\chi$, so that the initial evolution was in
the $\phi$ direction, until the field space path approached the valley
bottom at $\phi=0$.
In cases were $W_0$ is the dominant contribution to the
energy density, this evolution was shown
to lead to a positive spike in $\fnl$ with
its maximum scaling as\cite{EMST-11}
\be
\label{eq:valley_fnl}
\left. \fnl \right|_{\rm Max} \sim 0.3 \epsilon_*^{1/2} \frac{\Mpl}{\chi_*},
\ee
which was verified for a concrete model.\cite{EMST-11}
Again we clearly see the fine tuning
needed to generate a large
$\fnl$, and that for valleys a positive value of $\fnl$ is
expected, a feature expected to apply to generic valleys.

\section{Models} \label{sec:models}

The discussions of the preceding sections become invaluable
when we are confronted with a concrete model of inflation. By
identifying the ridge or the valley regions in the potential we
can employ the above results to understand
qualitatively how $\fnl$ evolves as these regions are
traversed, and whether a `natural' adiabatic limit will arise
at the end of the evolution.
Moreover, considering the conditions needed
for a large non-Gaussianity
at the adiabatic limit, and the
estimates for a transitory large non-Gaussianity,
we can identify initial conditions that give rise to
appreciable non-Gaussianities.
On the other hand, these analytic arguments can
only inform us so far. In the literature there are examples
of models for which analytic arguments suggest a large
non-Gaussianity once a natural focusing region is reached, but where
this only occurs after the slow-roll approximation ceases to accurately
describe the dynamics. Moreover, there are examples of models
for which no focusing region is available, but for which a large
non-Gaussianity is possible as inflation ends. The various
possibilities were classified in the introduction.
To study these cases fully numerical simulations are essential.

In our recent work\cite{EMST-11} we performed simulations
of a variety of models, using a numerical
implementation of the $\delta N$ formalism, confirming
the usefulness of the analytic estimates we have developed,
but also highlighting
how sensitive $\fnl$ can be to the exact time an adiabatic limit is
reached, i.e. before or after slow-roll ends.
In models for which no focusing region
exists, we also showed a strong sensitivity of $\fnl$ on the time scale of reheating.

Here we introduce
an additional model, which highlights the need
for numerical simulations, and also provides a new example of a model
with a large non-Gaussianity after natural focusing. In this case
the model can produce either a positive or negative asymptotic value
of $\fnl$ depending on the initial conditions.
We take a sum-separable
potential of the form
\be
W(\phi,\chi) = W_0+\frac{1}{2} m_\phi ^2 \phi^2 +
g \chi + \frac{1}{3} \lambda \chi ^3 + \frac{1}{4} \mu \chi^4
\label{eq:inflection}
\ee
where $W_0$, $m_\phi$, $g$, $\lambda$ and $\mu$ are all positive constants.
There is an inflection point in $ V(\chi)$ at
$\chi = 0$.
$W_0$ and $\mu$ are fixed by the requirement that there is a minimum with $W = 0$, at
$\chi_{\rm min} = -r$ where $r$ is taken to be positive. These
requirements impose two relations between the model parameters.
This model clearly belongs to the
class of potentials which contain a focusing region.

\begin{figure}[htb]
\center{\includegraphics[width = 11cm,height=9cm]{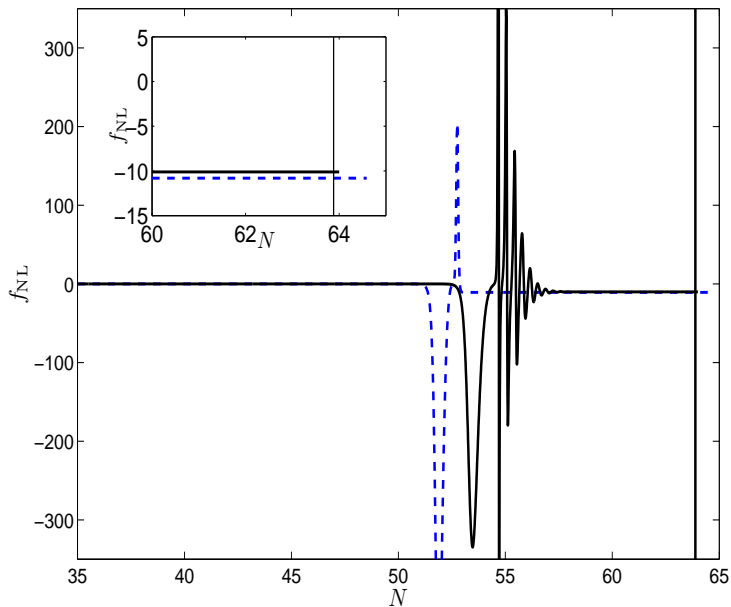}}
\caption{
Evolution of $\fnl$ for the potential \eqref{eq:inflection}
with parameter values and initial conditions
given in the text. The solid line shows the full numerical evolution
while the dashed line gives
the analytic solution.
The thin vertical line at $N \approx 64$ represents the analytically calculated end of
inflation when slow roll completely breaks down.
The fact that the adiabatic
limit value of $\fnl = -10.1$ (seen in the insert)
has been reached long before this point explains why the `HCA' analytic value of $\fnl -10.8$ is a good approximation. }
\label{fig:above_inflection}
\end{figure}

An explicit example is given by $g = (10^{-4}/0.18^2) m_\phi^2 \Mpl$, $\lambda = (100/0.18^2)
m_\phi^2 /\Mpl$ and $r = 0.14 \Mpl$, with the value of $m_\phi$ implicitly
chosen such that the value of $\zeta$ is normalised to be compatible with
CMB constraints\cite{WMAP7}.
Fig. \ref{fig:above_inflection} illustrates the evolution of $\fnl$ with the initial condition
$\phi^*=16\Mpl$, which supports close to $60$e-folds of inflation,
and $\chi^*= 0.0015 \Mpl$ which represents
an initial field position just above the inflection point.
The ridge--like shape of the inflection point
results in a negative spike in $\fnl$.
Subsequently the trajectories converge into a valley, while slow-roll is still maintained,
resulting in $\fnl$ temporarily becoming positive before eventually
tending to its limiting value. Recalling
the discussion of \S \ref{sec:analytics}, and
bearing in mind the initial conditions, we find that
$N_{,\chi} = V_\chi/V_\chi'$ is
the dominant first derivative of $N$, and
moreover $V_\chi'' /V_\chi$ is large and positive (since the
field is initially above the inflection point).
We therefore expect a large negative asymptotic value of $\fnl$,
which is borne out by the full numerical evolution. The numerical and HCA limiting values
are both shown in Fig. \ref{fig:above_inflection},
which show excellent agreement,
since slow-roll is maintained throughout the evolution.
We note that placing the field initially
at $\chi^*=-0.0015\Mpl$, where $V_\chi''$ is negative,
leads to nearly identical evolution, except that a positive $\fnl$ is ultimately reached (with the numerically calculated value of
$\fnl = 9.9$ in good agreement with the analytic HCA value of $\fnl=10.4$).
Finally, choosing initial
conditions very close to $\chi^*=0$, leads to a negligible $\fnl$,
which again is expected since $V_\chi'' \sim 0$.

\begin{figure}[htb]
\center{\includegraphics[width = 11cm, height=9cm]{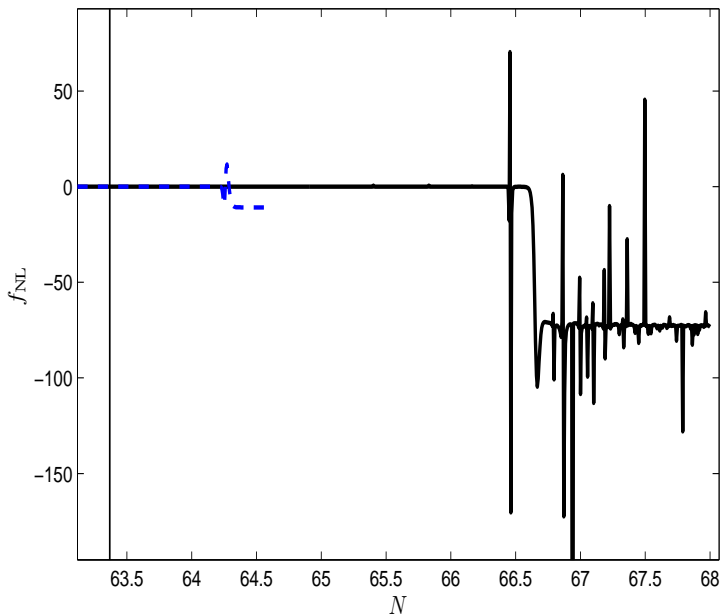}}
\caption{
Evolution of $\fnl$ for the potential \eqref{eq:inflection}
with parameters and initial conditions given in the text. As before, the solid line shows the full numerical evolution,
the dashed line the slow-roll
analytic evolution, and the thin vertical 
line shows the analytically calculated time when slow-roll
breaks down. A discrepancy between the analytic
and numerical evolution is not unexpected, though its
magnitude is perhaps a surprise.
}
\label{fig:inflection2}
\end{figure}

Now let us change the parameters so that instead of reaching the adiabatic limit
before inflation ends, it is reached just afterwards. We choose $g = 10^{-4} m_\phi^2\Mpl$, $\lambda = 100m_\phi^2/\Mpl$ and $r = 0.14 \Mpl$, and the same
initial conditions ($\phi^*=16\Mpl,\chi^*= 0.0015 \Mpl$).
It is clear that the analytic HCA value for $\fnl$ can no longer be
trusted, but one might hope that it gives at least
an indication of the true asymptotic value, particularly as the $\chi$ field will source a final
fraction of an e-fold of inflation as it rolls. Studying Fig.
\ref{fig:inflection2}, this is seen not to be the case, with the
analytic estimate (unchanged from the above case) being extremely inaccurate.

\section{Conclusions} \label{sec:conclusions}

We have studied the evolution of non-Gaussianity in multiple-field inflationary models,
using analytical and numerical methods.
We have shown that the descent of fields from a ridge
or their convergence into a valley can result in
significant growth of non-Gaussianity, with the two
cases being distinguished by the sign of $\fnl$.

To concretely predict non-Gaussianities one
can employ analytical expressions
or numerical methods. Currently, however,
the former rely on the slow-roll approximation
which limits their applicability. In such cases we
have demonstrated that numerical methods can be
invaluable.

To calculate observationally relevant values
of parameters such as $\fnl$, in practice it is
necessary that an adiabatic limit is reached
where these parameters become constant.
In order to determine the possible behaviours of
$\fnl$ as it reaches this limit, and
the techniques needed to calculate it accurately there,
we have found it is useful to
classify
inflationary models according to whether the adiabatic limit is reached naturally by the
convergence of field space trajectories
during the slow-roll regime, naturally
after slow-roll ends,
abruptly, or where no focusing region exists,
only after reheating occurs.
We have summarised the recent results from our paper\cite{EMST-11}
concerning these classes of models and
given a new illustration here using a new model.

Finally, we note that the new example
included in this work indicates that it is easy
to construct two-field
sum--separable models which exhibit
large $\fnl$ of positive or negative sign,
even when the adiabatic limit is reached naturally
during slow-roll inflation. We note, however, that
all such models exhibiting a large non-Gaussianity at
the adiabatic limit appear to
require a large degree of fine tuning
in their initial conditions.
\section*{Acknowledgements}
JE is supported by an Science and Technology Facilities Council Studentship. DJM
is supported by the Science and Technology Facilities Council grant ST/H002855/1. DS was
supported by the Science and Technology Facilities Council [grant numbers ST/F002858/1
and ST/I000976/1]. RT thanks the organisers of the Friedmann Seminar,
Rio de Janeiro, Brazil, for an enjoyable meeting.


\begin{thebibliography}{0} 
\bibitem{zetaAndZetaCon}
J.~M.~Bardeen, P.~J.~Steinhardt, M.~S.~Turner,
Phys.\ Rev.\ D {\bf 28}, 679 (1983);
D.~H.~Lyth,
Phys.\ Rev.\ D {\bf 31}, 1792 (1985);
D.~H.~Lyth, K.~A.~Malik, M.~Sasaki,
JCAP {\bf 0505}, 004 (2005);
G.~I.~Rigopoulos, E.~P.~S.~Shellard,
Phys.\ Rev.\ D {\bf 68}, 123518 (2003);
D.~Langlois and F.~Vernizzi,
Phys.\ Rev.\ D {\bf 72}, 103501 (2005);
S.~Weinberg,
Phys.\ Rev.\ D {\bf 70}, 043541 (2004).

\bibitem{Lyth}
D.~H.~Lyth,
Phys.\ Rev.\ D {\bf 31}, 1792 (1985).


\bibitem{Maldacena:2002vr}
J.~M.~Maldacena,
JHEP {\bf 0305}, 013 (2003).


\bibitem{GarciaBellido:1995qq}
J.~Garcia-Bellido, D.~Wands,
Phys.\ Rev.\ D {\bf 53}, 5437 (1996).


\bibitem{Gordon:2000hv}
C.~Gordon, D.~Wands, B.~A.~Bassett, R.~Maartens,
Phys.\ Rev.\ D {\bf 63}, 023506 (2001).




\bibitem{Vernizzi:2006ve}
F.~Vernizzi and D.~Wands,
JCAP {\bf 0605}, 019 (2006).

\bibitem{otherMethods}
G.~I.~Rigopoulos, E.~P.~S.~Shellard, B.~J.~W.~van Tent,
{\it Phys.\ Rev.\ D} {\bf 73}, 083521 (2006);
G.~I.~Rigopoulos, E.~P.~S.~Shellard, B.~J.~W.~van Tent,
{\it Phys.\ Rev.\ D} {\bf 76}, 083512 (2007);
S.~Yokoyama, T.~Suyama and T.~Tanaka,
{\it JCAP} {\bf 0707}, 013 (2007);
S.~Yokoyama, T.~Suyama and T.~Tanaka,
{\it Phys.\ Rev.\ D} {\bf 77}, 083511 (2008);
D.~J.~Mulryne, D.~Seery and D.~Wesley,
{\it JCAP} {\bf 1104}, 030 (2011);
D.~J.~Mulryne, D.~Seery and D.~Wesley,
{\it JCAP} {\bf 1001}, 024 (2010).

\bibitem{Seery:2005gb}
D.~Seery, J.~E.~Lidsey,
{\it JCAP} {\bf 0509}, 011 (2005);
D.~Seery, J.~E.~Lidsey and M.~S.~Sloth,
{\it JCAP} {\bf 0701}, 027 (2007);
D.~Seery, M.~S.~Sloth and F.~Vernizzi,
{\it JCAP} {\bf 0903}, 018 (2009).


\bibitem{hybrid}
L.~Alabidi,
{\it JCAP} {\bf 0610}, 015 (2006);
C.~T.~Byrnes, K.~Y.~Choi and L.~M.~H.~Hall,
{\it JCAP} {\bf 0902}, 017 (2009).


\bibitem{Byrnes:2008wi}
C.~T.~Byrnes, K.~Y.~Choi and L.~M.~H.~Hall,
{\it JCAP} {\bf 0810}, 008 (2008).


\bibitem{Kim:2010ud}
S.~A.~Kim, A.~R.~Liddle and D.~Seery,
{\it Phys.\ Rev.\ Lett.}\ {\bf 105}, 181302 (2010).



\bibitem{Peterson:2010mv}
C.~M.~Peterson and M.~Tegmark,
`Non-Gaussianity in Two-Field Inflation,'
[arXiv:1011.6675 [astro-ph.CO]].


\bibitem{EMST-11}
J.~Elliston, D.~J.~Mulryne, D.~Seery and R.~Tavakol,
`Evolution of $f_{NL}$ to the adiabatic limit,'
[arXiv:1106.2153 [astro-ph.CO]].



\bibitem{Starobinsky:1986fxa}
A.~A.~Starobinsky,
{\it JETP Lett.}\ {\bf 42}, 152 (1985).

\bibitem{Sasaki:1995aw}
M.~Sasaki and E.~D.~Stewart,
{\it Prog.\ Theor.\ Phys.}\ {\bf 95}, 71 (1996).


\bibitem{Wands:2000dp}
D.~Wands, K.~A.~Malik, D.~H.~Lyth and A.~R.~Liddle,
{\it Phys.\ Rev.\ D} {\bf 62}, 043527 (2000).


\bibitem{Lyth:2005fi}
D.~H.~Lyth and Y.~Rodriguez,
{\it Phys.\ Rev.\ Lett.}\ {\bf 95}, 121302 (2005).

\bibitem{trajectories}
S.~W.~Hawking,
{\it Astrophys.\ J.}\ {\bf 145}, 544 (1966);
A.~A.~Starobinsky,
Phys.\ Lett.\ B {\bf 117}, 175 (1982);
D.~S.~Salopek,
{\it Phys.\ Rev.\ D} {\bf 52}, 5563 (1995).




\bibitem{Battefeld:2006sz}
T.~Battefeld and R.~Easther,
{\it JCAP} {\bf 0703}, 020 (2007).


\bibitem{Choi:2007su}
K.~-Y.~Choi, L.~M.~H.~Hall and C.~van de Bruck,
{\it JCAP} {\bf 0702}, 029 (2007).


\bibitem{Wang:2010si}
T.~Wang,
{\it Phys.\ Rev.\ D} {\bf 82}, 123515 (2010).


\bibitem{WMAP7}
D.~Larson {\it et al.},
{\it Astrophys.\ J.\ Suppl.}\ {\bf 192}, 16 (2011)
[arXiv:1001.4635 [astro-ph.CO]];
E.~Komatsu {\it et al.} [WMAP Collaboration],
{\it Astrophys.\ J.\ Suppl.}\ {\bf 192}, 18 (2011).


\bibitem{Komatsu:2009kd}
E.~Komatsu {\it et al.},
`Non-Gaussianity as a Probe of the Physics of the Primordial Universe and
the Astrophysics of the Low Redshift Universe,'
[arXiv:0902.4759 [astro-ph.CO]].



\bibitem{HCA}
S.~A.~Kim, A.~R.~Liddle,
{\it Phys.\ Rev.\ D} {\bf 74}, 023513 (2006).
S.~A.~Kim, A.~R.~Liddle,
{\it Phys.\ Rev.\ D} {\bf 74}, 063522 (2006).



\end{thebibliography}
\end{document}